# In-plane to perpendicular magnetic anisotropy switching in heavily-Fe-doped ferromagnetic semiconductor (Ga,Fe)Sb with high Curie temperature

Shobhit Goel,[1, *] Le Duc Anh,[1,2,+] Nguyen Thanh Tu,[1,3] Shinobu Ohya,[1,2,4] and Masaaki Tanaka[1,4,†]

[1] *Department of Electrical Engineering and Information Systems, The University of Tokyo, 7-3-1 Hongo, Bunkyo-ku, Tokyo 113-8656, Japan*
[2] *Institute of Engineering Innovation, The University of Tokyo, 7-3-1 Hongo, Bunkyo-ku, Tokyo 113-8656, Japan*
[3] *Department of Physics, Ho Chi Minh City University of Pedagogy, 280, An Duong Vuong Street, District 5, Ho Chi Minh City 748242, Vietnam*
[4] *Center for Spintronics Research Network (CSRN), The University of Tokyo, 7-3-1 Hongo, Bunkyo-ku, Tokyo 113-8656, Japan*

We report switching of magnetic anisotropy (MA) from in-plane to perpendicular with increasing the thickness $d$ of a (001)-oriented ferromagnetic-semiconductor (FMS) (Ga$_{0.7}$,Fe$_{0.3}$)Sb layer with a high Curie temperature ($T_C > 320$ K), using ferromagnetic resonance at room temperature. We show that the total MA energy ($E_\perp$) along the [001] direction changes its sign from positive (in-plane) to negative (perpendicular) with increasing $d$ above an effective critical value $d_C^* \sim 42$ nm. We reveal that (Ga,Fe)Sb has two-fold symmetry in the film plane. Meanwhile, in the plane perpendicular to the film including the in-plane [110] axis, the two-fold symmetry with the easy magnetization axis along [110] changes to four-fold symmetry with easy magnetization axis along <001> with increasing $d$. This peculiar behavior is different from that of (Ga,Mn)As, in which only the in-plane MA depends on the film thickness and has four-fold symmetry due to its dominant cubic anisotropy along the <100> axes. This work provides an important guide for controlling the easy magnetization axis of high-$T_C$ FMS (Ga,Fe)Sb for room-temperature device applications.

*goel@cryst.t.u-tokyo.ac.jp
+anh@cryst.t.u-tokyo.ac.jp
†masaaki@ee.t.u-tokyo.ac.jp

# I. INTRODUCTION

Materials having perpendicular magnetic anisotropy (PMA) are very interesting compared with those having in-plane magnetic anisotropy (IMA) due to their ability to provide higher storage density and high thermal stability for information storage [1][2]. PMA is important for nonvolatile magnetic memory because PMA can reduce the critical current density for magnetization switching [3]–[5]. Also, PMA in ferromagnetic metal / semiconductor hybrid structures will be useful for semiconductor-based memory devices [6]; however, integrating metallic ferromagnets into semiconductor-based electronic devices, which is crucial for realizing low-power spin-based electronics, is challenging because of problems such as conductivity mismatch and magnetically dead layers at the metal/semiconductor interfaces.

On the other hand, ferromagnetic semiconductors (FMSs), which exhibit both magnetic and semiconducting properties, are more promising due to their good compatibility with the semiconductor technology. Over the past 20 years, most of the studies on FMSs have been focused on Mn-doped III-V *p*-type FMSs, such as (In,Mn)As and (Ga,Mn)As. Magnetic anisotropy (MA) of these Mn-doped III-V FMSs was observed and controlled [7]–[23]: The control of MA between IMA and PMA by strain, temperature, and carrier concentration change via an external gate voltage has been demonstrated. Moreover, it has been reported that the prototypical Mn-doped III-V FMS (Ga,Mn)As has complex properties of MA due to the co-existence of a cubic anisotropy field along the ⟨100⟩ axes, an uniaxial anisotropy fields along the ⟨110⟩ axes, and an additional uniaxial in-plane anisotropy field along the [100] axis. This issue is still a hot topic for researchers [23]. However, Mn-doped III-V FMSs such as (Ga,Mn)As are not suitable for practical device applications because of the following problems: (i) All the MA studies and device demonstrations using (Ga,Mn)As were performed at low temperature because of its low Curie temperature $T_C$ [the highest $T_C$ is 200 K for (Ga,Mn)As] [24]. (ii) All the Mn-doped FMSs are *p*-type (both *n*-type and *p*-type FMSs are necessary for realizing semiconductor-based spintronic devices).

Heavily Fe-doped narrow-gap III-V FMSs are promising alternatives to overcome the problems of the Mn-based FMSs. By using Fe as magnetic dopants, we can grow both *n*-type FMSs ((In,Fe)As [25][27], (In,Fe)Sb [28]) and *p*-type FMSs ((Al,Fe)Sb [29], (Ga,Fe)Sb [30]–[31]). This is because Fe atoms are mostly in the isoelectronic $Fe^{3+}$ state and do not supply carriers. Moreover, these Fe-doped FMSs, especially (Ga,Fe)Sb, are very promising for room-temperature device applications because of their high $T_C$ ($T_C$ > 400 K [32]). The observation of high-$T_C$ ferromagnetism in these Fe-doped FMSs is striking, because it is opposite to the prediction of the mean-field Zener model [33] which claims that $T_C$ will be high (low) in wide-gap (narrow-gap) FMSs. Therefore, by thoroughly investigating the magnetic properties, particularly the magnetic anisotropy, of these Fe-doped FMSs, it is expected to uncover various unknown aspects in the magnetic physics and materials science of FMSs. Previously by studying the MA of $(Ga_{0.8},Fe_{0.2})$Sb thin films (thickness $d$ = 15 nm), we found that IMA is dominant due to the large shape anisotropy constant ($K_{sh}$) [32]. Here in this work, we report a new finding that the switching of MA from IMA to PMA can be induced by increasing the thickness of (Ga,Fe)Sb over a critical value of 42 nm. This results provides an important guide for designing the MA of high-$T_C$ FMS (Ga,Fe)Sb thin films for device applications operating at room temperature.

## II. SAMPLE GROWTH AND CHARACTERIZATIONS

We grew a series of (001)-oriented $(Ga_{0.7},Fe_{0.3})$Sb thin films with various thicknesses $d$ of 15, 20, 30, 40, and 55 nm, which are named G15, G20, G30, G40, and G55, respectively. All the samples were grown by low-temperature molecular-beam epitaxy (LT-MBE) on an AlSb/AlAs/GaAs buffer layer using a semi-insulating GaAs (001) substrate [Fig. 1(a)]. The detailed growth procedure, crystal structure, X-ray diffraction (XRD), and other characterizations are presented in Supplementary Material (SM) [34]. From these detailed characterizations, we confirmed the intrinsic ferromagnetism of (Ga,Fe)Sb. XRD results show all the (Ga,Fe)Sb films with various thicknesses have tensile strain and almost the same value of epitaxial strain $\varepsilon \sim -1.58$ %, consistent with our previous report [32]. Therefore, in this study, we rule out any strain induced magnetic anisotropy change. We characterized the magnetic properties of all the samples using magnetic circular dichroism (MCD) spectroscopy and superconducting quantum interference device (SQUID) magnetometry. Figures 1(b)–1(e) show the magnetic-field $\mu_0 H$ dependence of MCD, which all show open hysteresis and confirms the room-temperature ferromagnetism in the four (Ga,Fe)Sb samples. In the MCD measurements, a magnetic field is always applied perpendicular to the film plane. Therefore, the MCD signal only reflects the perpendicular magnetization component. From Figs. 1(b)–1(e), we can see that by increasing the thickness $d$ from 20 to 55 nm, the magnetic field $\mu_0 H$ at which the MCD intensity saturates was reduced from 0.3 T to 0.1 T, and the coercive field increases from 40 mT to 135 mT. These changes in the MCD – $H$ curves thus clearly suggest that the PMA is enhanced with increasing the thickness of (Ga,Fe)Sb. The clear change in the magnetic anisotropy from IMA to PMA with increasing $d$ can also be confirmed in the magnetization hysteresis curves ($M – H$) measured at 10 K using SQUID, as shown in Figs. 2(a) and 2(b).

## III. FERROMAGNETIC-RESONANCE (FMR) MEASUREMENTS AND THEORETICAL MODEL

We used ferromagnetic resonance (FMR) to estimate the MA constants of the $(Ga_{0.7},Fe_{0.3})Sb$ thin films in a similar fashion as reported in our previous work [32]. The samples were placed at the center of a $TE_{011}$ microwave cavity, at which the rf magnetic field and rf electric field of the microwave are largest and smallest, respectively. The microwave frequency used for the measurement is 9.07 GHz (see SM [34]). All the samples were measured under a microwave power $P$ = 200 mW at 300 K. In all the samples, we observed clear FMR signals from the (Ga,Fe)Sb thin films at room temperature. Figure 3(a) shows FMR spectra of the (Ga,Fe)Sb films with the magnetic field $H$ // [110] (red circles) and $H$ // [001] (black squares). We note that a raw FMR spectrum included a background signal, which was separately detected by measuring the FMR spectrum without a sample in the cavity and then was subtracted from the raw data. The crossing point of the raw FMR spectrum and the background FMR spectrum is the resonance field $\mu_0 H_R$ [violet arrows in Fig. 3(a)] (see SM [34]). We also measured the dependence of $\mu_0 H_R$ on the magnetic-field direction, when $H$ is rotated in the film plane between the [110] and [$\bar{1}$10] axes and rotated in the plane between the direction normal to the film plane ($H$ // [001]) and the in-plane direction ($H$ // [110]), as shown in the Figs. 3(b) and 3(c), respectively. Figure 1(f) shows the coordinate system in the FMR measurements, where the direction of $H$ is changed between perpendicular and in-plane directions. Here, $\theta_H$ and $\theta_M$ are the angles of $H$ and the magnetization $M$ from the [001] direction, and $\varphi_H$ and $\varphi_M$ are the angles of $H$ and $M$ from the [100] direction in the film plane, respectively. We found that $\mu_0 H_R$ of the FMR spectra strongly depends on $d$. For G15–G40 ($d$ = 15 – 40 nm), $\mu_0 H_R$ is smaller when measured with $H$ // [110] than that with $H$ // [001], meaning that the thin (Ga,Fe)Sb samples have IMA. The difference of $\mu_0 H_R$ between $H$ // [110] and $H$ // [001] decreases with increasing $d$ up to 40 nm. Note that at $d$ = 40 nm (sample G40), the difference between $\mu_0 H_R$ of the FMR spectra with $H$ // [110] and $H$ // [001] is closest to zero (~30 mT). When *d* is increased to 55 nm (sample G55), $\mu_0 H_R$ measured with **H** // [001] becomes smaller than that with **H** // [110], which indicates that the easy magnetization axis is perpendicular to the film plane in G55. These results are consistent with the SQUID results at 300 K. We also note that the FMR spectrum of G55 has two FMR peaks when **H** // [001]. The origin may be due to the inhomogeneity of the Fe-atom distribution [see Fig. 5(c)], which causes nonuniform excitations of the magnetization precession, resulting in an extra resonance mode [34].

Next, the detailed angular dependence of $\mu_0 H_R$ on the **H** direction ($\theta_H$, $\varphi_H$) of all the samples is represented as the solid circles in Figs. 3(b) and 3(c). In Fig. 2(b), when rotating **H** in the perpendicular plane, for G15, G20, and G30, we observed two-fold symmetry, where $\mu_0 H_R$ decreased smoothly with increasing $\theta_H$ from 0° (**H** // [001]) to 90° (**H** // [110]). For G40, $\mu_0 H_R$ first increases and then decreases, which means that four-fold symmetry appears and it becomes clearer in G55, where $\mu_0 H_R$ is smaller at $\theta_H = 0°$ (**H** // [001]) than at $\theta_H = 90°$ (**H** // [110]). In Fig. 3(c), when **H** is rotated in the film plane, we observed only two-fold symmetry, where the $\mu_0 H_R$ value decreases smoothly with increasing $\varphi_H$ from 45° (**H** // [110]) to 135° (**H** // [$\bar{1}$10]).

To determine the MA constants, we used the free energy density $E_{\theta_M, \varphi_M}$ given by [35]

$$E_{\theta_M,\varphi_M} = \frac{1}{2}\mu_0 M_S \left\{ -2H[\cos\theta_M \cos\theta_H + \sin\theta_M \sin\theta_H \cos(\varphi_M - \varphi_H)] + M_S \cos^2\theta_M - H_{2\perp}\cos^2\theta_M \right. $$
$$\left. - \frac{H_{4\perp}}{2}\cos^4\theta_M - H_{2//}\sin^2\theta_M \sin^2\left(\varphi_M - \frac{\pi}{4}\right) \right\}. \qquad (1)$$

Here in { }, the first term describes the Zeeman energy, the second term is the demagnetizing energy or shape anisotropy, and the other terms are the MA energies, where $\mu_0 H_{2\perp}$ and $\mu_0 H_{4\perp}$ are the perpendicular uniaxial and cubic anisotropy fields, respectively, and $\mu_0 H_{2//}$ is the in-plane uniaxial anisotropy field. $\mu_0 H_{2\perp}$, $\mu_0 H_{2//}$, and $\mu_0 H_{4\perp}$ are defined as $\mu_0 H_{2\perp} = 2K_{2\perp}/M_S$, $\mu_0 H_{2//} = 2K_{2//}/M_S$, and $\mu_0 H_{4\perp} = 2K_{4\perp}/M_S$, where $K_{2\perp}$, $K_{4\perp}$, and $K_{2//}$ are the corresponding perpendicular uniaxial, cubic, and in-plane uniaxial MA constants, respectively. We also note that the sign of the MA constants can be determined by minimizing Eq. (1) with respect to $\theta_M$ and $\varphi_M$. $K_{2\perp}$ and $K_{4\perp}$ are positive when the easy magnetization axis is along [001] and negative when it is along [110]. On the other hand, $K_{2//}$ is positive when it is along [$\bar{1}$10] and negative when it is along [110]. The condition at $\varphi_M = \frac{\pi}{4}$ and $\theta_M = 0$, i.e. in the plane containing the [110] and [001] axis, along with $\mu_0 H = 0$, corresponds to the MA switching between IMA and PMA at $d = d_C^*$ or $d = d_C$, where $d_C^*$ is an effective critical thickness with $K_{sh}$ and $d_C$ is a critical thickness without $K_{sh}$. The corresponding magnetic anisotropy energy (MAE) equation is given by

$$E_{\theta_M=0,\, \varphi_M=\frac{\pi}{4}} = -\frac{1}{2}M_S\mu_0 H_{2\perp} - \frac{1}{4}M_S\mu_0 H_{4\perp} + \frac{1}{2}\mu_0 M_S^2$$

$$E_{\theta_M=0,\, \varphi_M=\frac{\pi}{4}} = E_\perp = E_\perp^0 + E_{sh} = -\left(K_{2\perp} + \frac{K_{4\perp}}{2}\right) - K_{sh}. \qquad (2)$$

where $E_\perp^0 = -\left(K_{2\perp} + \frac{K_{4\perp}}{2}\right)$ is the perpendicular MAE, $E_{sh} = -K_{sh}\, (= \frac{1}{2}\mu_0 M_S^2)$ is the shape anisotropy energy, and $E_\perp$ is the total MAE including $E_{sh}$ in the [001] direction (i.e. $\theta_M = 0$). We note that the MA energies ($E_\perp$ and $E_\perp^0$) are opposite in sign to the MA constants ($K_{2\perp}$ and $K_{4\perp}$). The film shows IMA when $E_\perp > 0$ and PMA when $E_\perp < 0$ according to Eq. (2). Using Eq. (1) along with the well-known Landau-Lifshitz-Gilbert (LLG) equation [36],[37] and Smith-Beljers expression [38], we derived the following equation for the rotation of $\boldsymbol{H}$ in the perpendicular plane ($\varphi_H = \varphi_M = 45°$), i.e. in the ($\bar{1}$10) plane (see SM [34]);

$$\left(\frac{\omega}{\gamma}\right)^2 = \mu_0^2 [H_R \cos(\theta_H - \theta_M) + (-M_S + H_{2\perp})\cos^2\theta_M + H_{4\perp}\cos^4\theta_M - H_{2//}]$$

$$\times \left[H_R \cos(\theta_H - \theta_M) + \left(-M_S + H_{2\perp} + \frac{H_{4\perp}}{2}\right)\cos 2\theta_M + \frac{H_{4\perp}}{2}\cos 4\theta_M\right]. \quad (3)$$

Here, $\omega$ is the Larmor angular frequency, $\gamma = g\mu_B/\hbar$ is the gyromagnetic ratio, where $g$, $\mu_B$, and $\hbar$ are the $g$-factor, Bohr magneton, and reduced Planck's constant, respectively, and $\mu_0 H_R$ is the resonance field. Similarly, the equation for the rotation of **H** in the film plane ($\theta_H = \theta_M = 90°$), i.e. in the (001) plane, is given by

$$\left(\frac{\omega}{\gamma}\right)^2 = \mu_0^2 \left[H_R \cos(\varphi_H - \varphi_M) + M_S - H_{2\perp} + H_{2//}\sin^2\left(\varphi_M - \frac{\pi}{4}\right)\right]$$

$$\times \left[H_R \cos(\varphi_H - \varphi_M) - H_{2//}\cos\left(2\varphi_M - \frac{\pi}{2}\right)\right]. \quad (4)$$

We fit Eqs. (3) and (4) to the experimental FMR data of Figs. 2(b) and 2(c), respectively, as will be described later. We used $\mu_0 H_{2\perp}$, $\mu_0 H_{4\perp}$, $\mu_0 H_{2//}$, and $\gamma$ (or $g$-factor) as fitting parameters. Using the obtained fitting values of the MA fields ($\mu_0 H_{2\perp}$, $\mu_0 H_{4\perp}$, and $\mu_0 H_{2//}$), we first estimated the MA constants ($K_{2\perp}$, $K_{4\perp}$, and $K_{2//}$), and then we obtained the value of $K_{sh}$ ($= -\frac{1}{2}\mu_0 M_S^2$) by using the saturation magnetization ($M_s$) values obtained from the SQUID measurements. Finally, $E_\perp^0$ ($= -K_{2\perp} - \frac{K_{4\perp}}{2}$) and $E_\perp$ ($= -K_{2\perp} - \frac{K_{4\perp}}{2} - K_{sh}$) were estimated for all the samples.

## IV. RESULTS AND DISCUSSIONS

Figures 3(b) and 3(c) show the fitting curves (see the black solid curves) derived from Eqs. (3) and (4), which reproduce the observed angular dependence of the FMR fields very well for all the samples. Table I shows the derived fitting parameters at room temperature (300 K). In Table I, we note that $\mu_0 H_{2\perp}$ rapidly increases as $d$ increases, while $\mu_0 H_{2//}$ is almost constant. We also note that in G40 and G55, $\mu_0 H_{4\perp}$ appears and increases with increasing $d$. These findings show that there is a strong $d$ dependence of magnetocrystalline MA in the (Ga,Fe)Sb films. In Table I, we also note that the estimated $g$ factor values show deviation from 2. This deviation in $g$ from 2 is due to the spin orbit interaction [39], which indicates that Fe atoms in the (Ga,Fe)Sb thin films are not only in the 3+ but also in the 2+ state. Figure 4(a) summarizes the obtained magnetocrystalline MA constants ($K_{2\perp}$, $K_{4\perp}$, and $K_{2//}$) as a function of $d$. $K_{2\perp}$ and $K_{4\perp}$, which have positive signs, align the magnetization normal to the film plane. In the film plane, $K_{2//}$ has a positive sign, which means that the in-plane easy magnetization axis of (Ga,Fe)Sb is along [$\bar{1}$10]. As shown in Fig. 4(a), $K_{2\perp}$ and $K_{4\perp}$ increase with increasing $d$, which causes the sign change of $E_\perp^0$ from positive (IMA) to negative (PMA) at $d = d_C \sim 17$ nm [see the violet rhombuses in Fig. 4(c)]. To understand the $d$ dependence of $E_\perp^0$, we separated $E_\perp^0$ using

$$E_\perp^0 = E_{\perp,1}^0(d) + E_{\perp,0}^0. \quad (5)$$

Here, $E_{\perp,1}^0(d) = ad$ is the thickness-dependent component, where $a$ is a coefficient, and $E_{\perp,0}^0$ is the thickness-independent component. The dashed line in Fig. 4(c) shows the fitting carried out using Eq. (5), by which we obtained a negative value of $a = -1.49 \times 10^{11}$ J/m$^2$ and a positive value of $E_{\perp,0}^0 = 2.5$ kJ/m$^3$. The negative value of $E_{\perp,1}^0(d)$, which increases with $d$, causes PMA and the positive value of $E_{\perp,0}^0$ causes IMA in our (Ga,Fe)Sb films.

The critical thickness $d_C \sim 17$ nm where $E_\perp^0$ changes sign is smaller than the thickness value of $\sim 40$ nm where the easy magnetization axis switches from the in-plane to perpendicular-to-plane direction as revealed by the FMR measurements in Fig. 3. This is due to the additional contribution from the shape anisotropy energy $E_{sh}$ of the (Ga,Fe)Sb thin films, which tends to align the magnetization in the film plane. Fig. 4(b) shows the saturation magnetization $M_S$ (magnetic moment per unit volume) of all the (Ga,Fe)Sb films as a function of $d$. The magnetic moment per unit volume should be independent of the film thickness for homogenous films. Thus, if the (Ga,Fe)Sb films are homogeneous, then $M_S$ should be independent of $d$. However, $M_S$ increases as $d$ increases as shown in Fig. 4(b), suggesting that inhomogeneity appears in the thick (Ga,Fe)Sb films. To understand its origin, we characterize the microscopic structure of the 40-nm-thick $(Ga_{0.7},Fe_{0.3})Sb$ sample (G40). Figure 5(a) shows the high-resolution scanning transmission electron microscopy (STEM) lattice image projected along the [110] axis, and Fig. 5(b) is the magnified image of the area indicated by the red rectangle in Fig. 5(a). These results, together with the transmission electron diffraction (TED) image in Fig. 5(c), clearly indicate the zinc-blende crystal structure of (Ga,Fe)Sb. However, there are regions with different contrast, which suggests that there are fluctuations in the local Fe density in the (Ga,Fe)Sb layer induced by spinodal decomposition. From the energy-dispersive X-ray spectroscopy (EDX) mapping of Fe atoms [Fig. 5(d)], it can be seen that there are nano-columnar-like Fe-rich regions growing from the interface [shown by the white arrows in Figs. 5(a), 5(b), and 5(d)]. These zinc-blende Fe-rich nano-columns [relatively dark regions in Figs. 5(a) and 5(b)] are elongated along the growth axis [001] with an in-plane diameter of several nm. We infer that the formation of Fe-rich nanocolumnar regions causes the increase of $M_S$ with $d$, which also complicates the estimation of the shape anisotropy $K_{sh}$ in the thick (Ga,Fe)Sb samples. For the sake of simplicity, we assume that there is no nanocolumars in the thinnest sample G15 and use the estimated value $K_{sh}$ of G15, which is –3.6 kJ/m$^3$, for the $K_{sh}$ of all the other (Ga,Fe)Sb samples. As shown by the red circles in Fig. 4(c), the value of $E_\perp$ (= $E_\perp^0$ + $E_{sh}$) changes sign, corresponding to a switching between IMA and PMA, at an effective critical thickness $d_C^*$ ~ 42 nm. The $d_C^*$ value agrees well with the experimental observation by FMR in Fig. 3, which thus suggests the validity of our simple assumption of the shape anisotropy. Previously, the observation of a change between IMA and PMA with changing the film thickness was reported in (In,Mn)As, where the MA switching was induced with a μm-scale difference in thickness and at very low temperature (4.2 K) [40]. Thus far, (Ga,Fe)Sb is the only FMS that can show room-temperature switching of the easy magnetization axis from in-plane to perpendicular just by changing the thickness in the nm-scale, which is a very simple way to control MA.

We suggest that the $K_{2\perp}$ component, which induces PMA in the thicker (Ga,Fe)Sb samples, is caused by the shape anisotropy of the columnar-like Fe-rich regions, whose shape is elongated along the [001] axis (growth direction). These Fe rich regions appear to be increasing with $d$, which in turn increases the perpendicular shape anisotropy, and thus $K_{2\perp}$ strongly depends on $d$. Similarly, in the thick G40 and G50 samples, the perpendicular four-fold cubic anisotropy ($K_{4\perp}$ component) also appears when $H$ is rotated between the [001] and [110] axes [Fig. 4(a)]. This thickness-dependence of $K_{4\perp}$ suggests that its origin should be related to the magnetocrystalline anisotropy of the Fe-rich nanocolumns formed in the thick (Ga,Fe)Sb sample. This origin, however, needs further careful investigations to be clarified. We also note that in (Ga,Fe)Sb with the Fe concentration lower than 20%, PMA does not appear due to the absence of Fe rich regions [30]. On the other hand, the weak in-plane two-fold symmetry along the [$\bar{1}$10] axis ($K_{2//}$ component) is observed in all the samples. This may be explained by the band structure of heavily-Fe-doped (Ga,Fe)Sb, where the Fermi level ($E_F$) is located in the Fe-related impurity band (IB) in the band gap [41]. Like (Ga,Mn)As, where Mn dimers have preferable orientation along the [$\bar{1}$10] axis [42],[43], we suggest that in (Ga,Fe)Sb there is a preferable Fe distribution along the [$\bar{1}$10] axis, thereby making the [$\bar{1}$10] axis an easy magnetization axis. However, unlike (Ga,Mn)As [15],[20], there is no four-fold symmetry in the film plane along the <100> axes in the case of (Ga,Fe)Sb. In (Ga,Mn)As with low Mn-concentrations, the position of $E_F$ in the Mn-related IB is very close to the VB maximum. Therefore, the symmetry of VB [light-hole and heavy-hole bands] strongly affects the MA via the *p-d* exchange interaction and causes strong in-plane cubic anisotropy [44],[45].

## V. CONCLUSION

We have revealed the strong *d* dependence of MA in (Ga,Fe)Sb using FMR measurements at 300 K. In the thin (Ga,Fe)Sb samples (G15, G20, G30, and G40), IMA is observed. This result is consistent with our previous report, where the shape anisotropy is dominant over the perpendicular magnetocrystalline anisotropy [32]. However, when *d* is large (G55), the PMA is observed, which is possibly caused by the formation of the nano-columnar-like Fe-rich regions due to spinodal decomposition. These findings suggest that the local fluctuation of Fe density in heavily-Fe-doped (Ga,Fe)Sb plays an important role in determining the magnetic properties of these thin films, and that the easy magnetization axis of (Ga$_{0.7}$,Fe$_{0.3}$)Sb can be controlled by changing the film thickness. The observation of FMR at room temperature and the control of MA are important steps towards understanding the magnetic properties and device applications of (Ga,Fe)Sb.

## ACKNOWLEDGMENTS

This work was partly supported by Grants-in-Aid for Scientific Research by MEXT (No. 26249039, No. 17H04922, No. 16H02095, and 18H03860), CREST Program (JPMJCR1777) of JST, the Spintronics Research Network of Japan (Spin-RNJ), and the Murata Science Foundation.

TABLE I. Properties of the $(Ga_{0.7},Fe_{0.3})Sb$ samples with various thicknesses ($d$) studied in this work at room temperature (300 K): saturation magnetization ($M_S$) measured by SQUID, magnetic anisotropy fields; perpendicular uniaxial anisotropy field ($\mu_0 H_{2\perp}$), cubic anisotropy field ($\mu_0 H_{4\perp}$), in-plane uniaxial anisotropy field ($\mu_0 H_{2//}$), and $g$ factor obtained by the fitting to the $\boldsymbol{H}$ angle dependence of the resonant field.

| Sample | $d$ (nm) | $M_S \times 10^4$ (A/m) | $\mu_0 H_{2\perp}$ (mT) | $\mu_0 H_{4\perp}$ (mT) | $\mu_0 H_{2//}$ (mT) | $g$ factor |
|---|---|---|---|---|---|---|
| G15 | 15 | 7.58 | $-10.7 \pm 0.2$ | – | $10 \pm 0.2$ | $2.04 \pm 0.01$ |
| G20 | 20 | 7.65 | $18.6 \pm 0.3$ | – | $4.8 \pm 0.3$ | $2.05 \pm 0.02$ |
| G30 | 30 | 8.05 | $59 \pm 0.1$ | – | $13 \pm 0.1$ | $2.07 \pm 0.04$ |
| G40 | 40 | 8.42 | $69.8 \pm 0.2$ | $31 \pm 0.2$ | $8 \pm 0.2$ | $2.06 \pm 0.02$ |
| G55 | 55 | 9.20 | $94 \pm 0.2$ | $42 \pm 0.2$ | $11 \pm 0.2$ | $2.08 \pm 0.04$ |

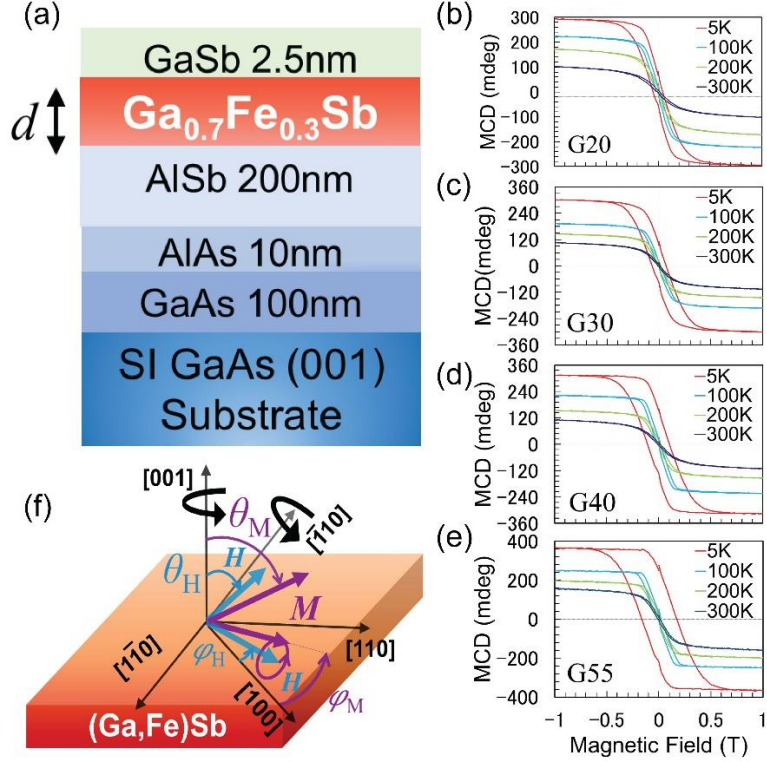

FIG. 1. (a) Schematic illustration of the (001)-oriented ($Ga_{0.7}$,$Fe_{0.3}$)Sb sample with various thicknesses $d$ = 15 nm, 20 nm, 30 nm, 40 nm, and 55 nm (G15, G20, G30, G40, and G55) grown on a semi-insulating GaAs (001) substrate. (b)–(e) MCD – $H$ curves of (Ga,Fe)Sb samples (G20, G30, G40, and G55) at different temperatures. The (Ga,Fe)Sb thin films in all the samples exhibit clear ferromagnetism with $T_C$ > 320 K. (f) Coordinate system used in the ferromagnetic resonance (FMR) measurement. The sample is rotated about the [$\bar{1}$10] and [001] axes. $\theta_H$ and $\theta_M$ are the angles of the applied magnetic field $H$ and magnetization $M$ from the [001] axis, and $\varphi_H$ and $\varphi_M$ are the angles of $H$ and $M$ from the [100] axis in the film plane.

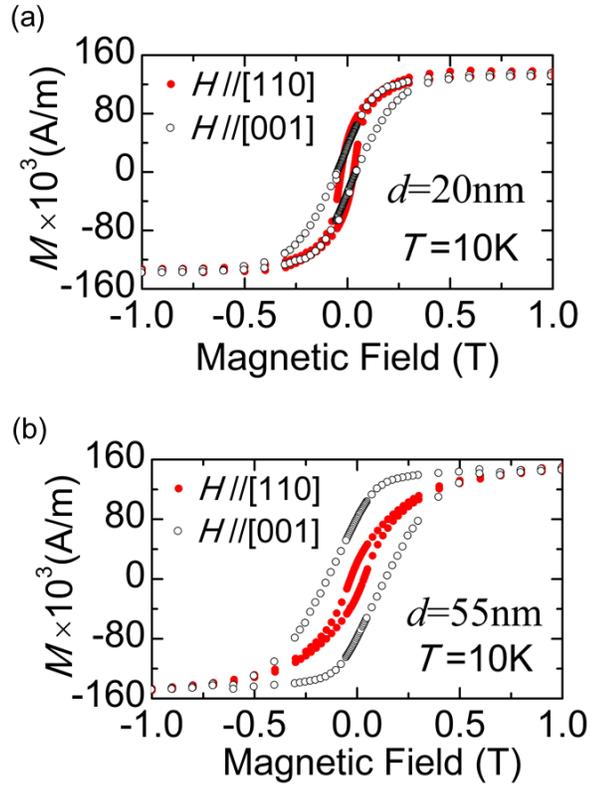

FIG. 2. Magnetization hysteresis curves ($M - H$) measured at 10 K for $(Ga_{0.7},Fe_{0.3})Sb$ with different thicknesses (a) $d = 20$ nm, and (b) $d = 55$ nm, when a magnetic field $\boldsymbol{H}$ was applied in the film plane along the [110] axis (red circles) and perpendicular to the plane along the [001] axis (black circles).

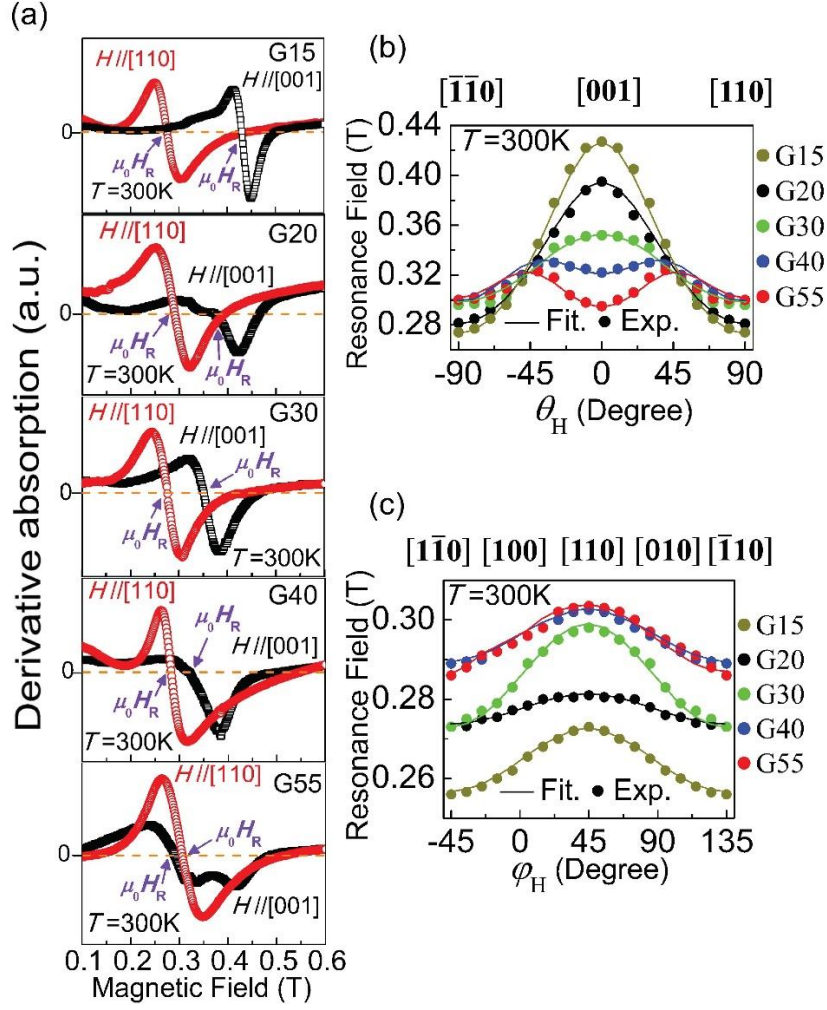

FIG. 3. (a) FMR spectra of the $(Ga_{0.7},Fe_{0.3})Sb$ samples at room temperature (300 K) when the magnetic field $\boldsymbol{H}$ was applied along the in-plane [110] axis (red circles) and the perpendicular [001] axis (black squares). The resonance field $\mu_0 H_R$ values are marked by violet arrows. (b) $\mu_0 H_R$ (solid symbols) as a function of the angle $\theta_H$ at 300 K. The solid line is a fitted curve by Eq. (3). (c) $\mu_0 H_R$ (solid symbols) as a function of the angle $\varphi_H$ at 300 K. The solid line is a fitted curve by Eq. (4).

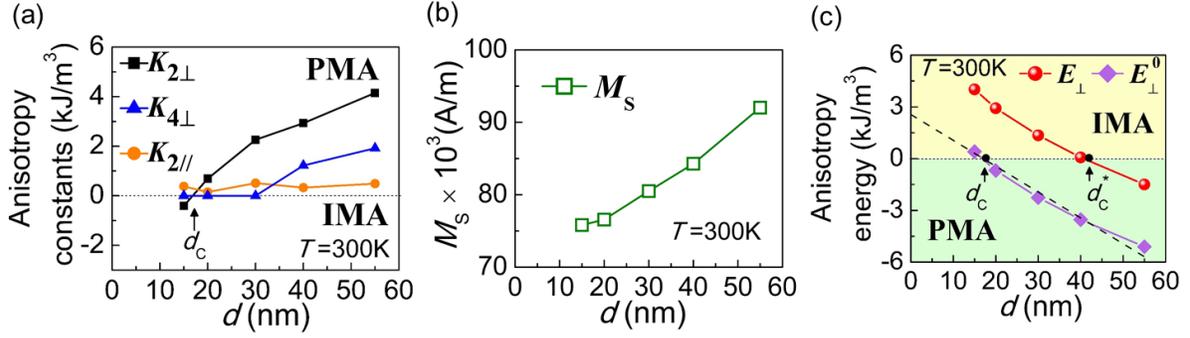

FIG. 4. Thickness ($d$) dependence of the (a) MA constants; the perpendicular uniaxial anisotropy constant $K_{2\perp}$ (squares), the cubic anisotropy constant $K_{4\perp}$ (triangles), the in-plane uniaxial anisotropy constant $K_{2//}$ (circles), (b) the saturation magnetization $M_S$ (squares), and (c) the perpendicular magnetocrystalline anisotropy energy $E_\perp^0$ (violet rhombuses) and the total MAE $E_\perp$ (red circles) of (Ga,Fe)Sb along [001]. The black dashed line is the fitting line for $E_\perp$ obtained using Eq. (5).

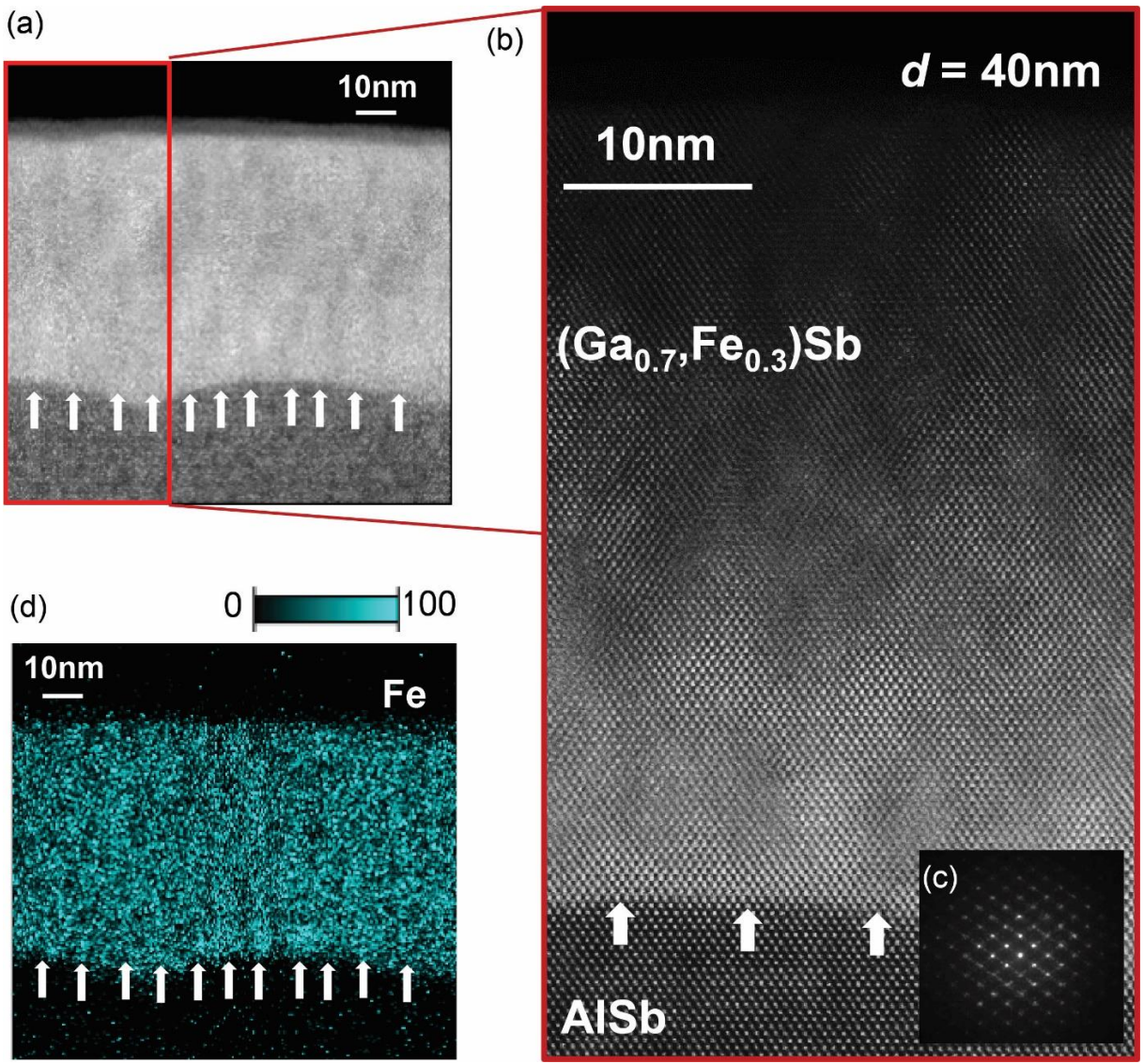

FIG. 5. (a) Cross-sectional scanning transmission electron microscopy (STEM) lattice image. (b) High-resolution STEM lattice image taken in the area marked by the red rectangle in (a). (c) Transmission electron diffraction (TED) pattern. (d) Energy dispersive X-ray analysis (EDX) of the Fe distribution in the 40-nm-thick $(Ga_{0.7},Fe_{0.3})Sb$ (G40), where nano-columnar-like Fe rich regions are shown by the white arrows.

# Supplementary Material
# In-plane to perpendicular magnetic anisotropy switching in heavily-Fe-doped ferromagnetic semiconductor (Ga,Fe)Sb with high Curie temperature

Shobhit Goel,[1] Le Duc Anh,[1,2] Nguyen Thanh Tu,[1,3] Shinobu Ohya,[1,2,4] and Masaaki Tanaka[1,4]

[1]*Department of Electrical Engineering and Information Systems, The University of Tokyo, 7-3-1 Hongo, Bunkyo-ku, Tokyo 113-8656, Japan*
[2]*Institute of Engineering Innovation, The University of Tokyo, 7-3-1 Hongo, Bunkyo-ku, Tokyo 113-8656, Japan*
[3]*Department of Physics, Ho Chi Minh City University of Pedagogy, 280, An Duong Vuong Street, District 5, Ho Chi Minh City 748242, Vietnam*
[4]*Center for Spintronics Research Network (CSRN), The University of Tokyo, 7-3-1 Hongo, Bunkyo-ku, Tokyo 113-8656, Japan*

1. **Sample growth and crystal structure of $(Ga_{0.7},Fe_{0.3})Sb$**

We grew four samples with different thicknesses of a *p*-type FMS $(Ga_{1-x},Fe_x)Sb$ layer ($x$ = 0.3, $d$ = 20, 30, 40, and 55 nm) via low-temperature molecular-beam epitaxy (LT-MBE) on AlSb/AlAs/GaAs using semi-insulating GaAs (001) substrates. The schematic structure of our samples is shown in Fig. 1(a) of the main text and in Fig. S1(a). In all the samples, after growing a 100-nm-thick GaAs layer on a semi-insulating GaAs substrate at a substrate temperature $T_S$ = 550 °C, we grew a 10-nm-thick AlAs layer at the same $T_S$. Then, we grew a 200-nm-thick AlSb layer at $T_S$ = 470 °C. After that, a $(Ga_{1-x},Fe_x)Sb$ layer with an Fe concentration $x$ = 30 % and a thickness $d$ = 20 – 55 nm was grown with a growth rate of 0.5 µm/h and an $Sb_4$ pressure of 7.8 – 8×10$^{-5}$ Pa (the best growth condition for the growth of (Ga,Fe)Sb) at $T_S$ = 250 °C for all the samples. Finally, we grew a 2.5-nm-thick GaSb cap layer in all the samples to avoid surface oxidation of the (Ga,Fe)Sb layer. As shown in Fig. S1(b), the in-situ reflection high-energy electron diffraction (RHEED) patterns in the [1̄10] azimuth of the (Ga,Fe)Sb thin films in all the four samples are bright and streaky with (1×3) reconstruction, thereby indicating good two-dimensional growth of a zinc-blende crystal structure.

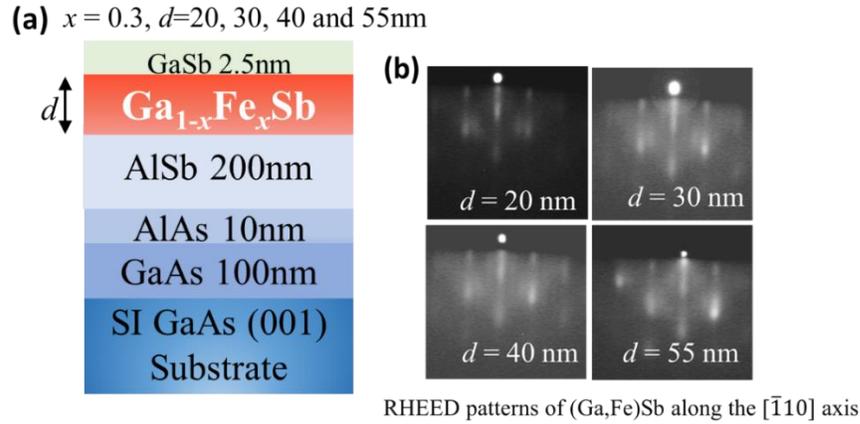

Figure S1. (a) Schematic illustration of the sample structure studied in this work. (b) In-situ reflection high-energy diffraction (RHEED) patterns taken along the [1̄10] axis during the MBE growth of $(Ga_{0.7},Fe_{0.3})Sb$ with various thicknesses d = 20 nm, 30 nm, 40 nm, and 55 nm.

Figure S2 shows the microstructure analyses [high resolution scanning transmission electron microscopy (STEM), transmission electron diffraction (TED) patterns]. On the right side of Fig. S2 are the TED patterns of point#1, point#2, and point#3 marked in the STEM image, where point#1 and point#2 correspond to the Fe-rich regions and point#3 corresponds to the (Ga,Fe)Sb matrix. From the TED patterns, we can see that although the crystal structures at point#1 and point#2 contain twin defects, all the three points have the same zinc-blende crystal structure. There is no second phase or precipitation. From these structural and other characterizations including MCD measurements (see the next section), we concluded that $(Ga_{0.7},Fe_{0.3})Sb$ has intrinsic ferromagnetism.

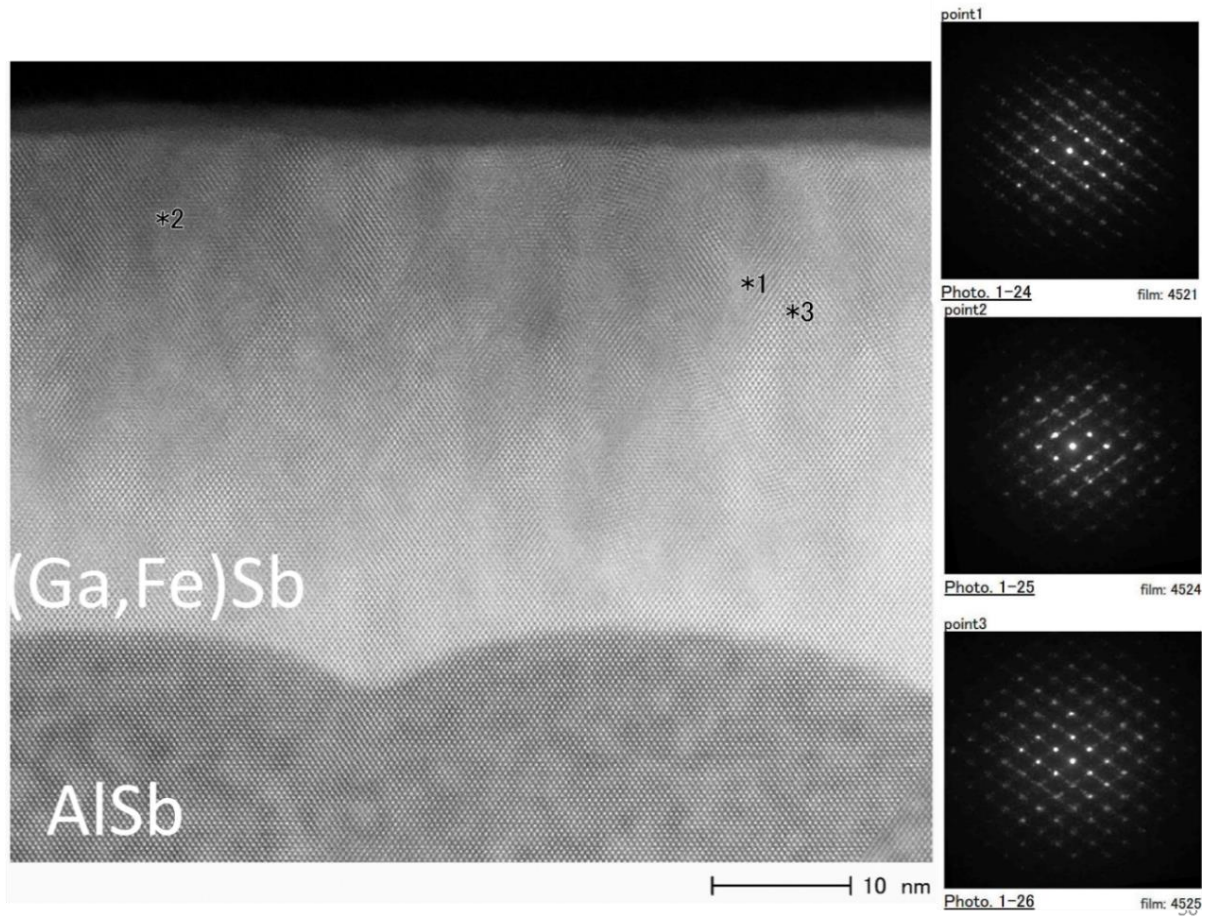

Figure S2. Scanning transmission electron microscopy (STEM) lattice image and transmission electron diffraction (TED) patterns of a 40 nm-thick $(Ga_{0.7},Fe_{0.3})Sb$ film grown on a AlSb buffer layer

2. **Magnetic circular dichroism (MCD) spectroscopy, Arrott plot and**

   In this work, we investigate the magneto-optical properties of the $(Ga_{0.7},Fe_{0.3})Sb$ samples by magnetic circular dichroism (MCD) spectroscopy in a reflection configuration. The reflection MCD intensity is given by the difference in the optical reflectivity for right ($R_{\sigma+}$) and left ($R_{\sigma-}$) circular polarization of light. The MCD intensity is expressed as

$$MCD = \frac{90}{\pi}\frac{(R_{\sigma+} - R_{\sigma-})}{2R} = \frac{90}{\pi}\frac{1}{2R}\frac{dR}{dE}\Delta E ,$$ where $R$ is the optical reflectivity, $E$ is the photon energy, and $\Delta E$ is the Zeeman splitting energy. Since, MCD is proportional to $dR/dE$ and $\Delta E$, it directly probes the spin-polarized band structure of the measured material. Thus, MCD is a powerful tool to characterize the intrinsic magnetic properties of (Ga,Fe)Sb [30]. Figures 1(b)–1(e) in the main manuscript show the MCD – $H$ characteristics of all the samples ($d$ = 20, 30, 40, and 55 nm) measured at different temperatures. Figure S3 show the MCD spectra of the 40 nm-thick $(Ga_{0.7},Fe_{0.3})Sb$ film measured at 5 K with a magnetic field of 0.2 T, 0.5 T and 1 T normalized by the intensity of $E_1$ peak at 1 T. MCD spectra have the $E_1$ peak around ~ 2.3 eV, reflecting the band structure of zinc-blende (Ga,Fe)Sb. As shown in Fig. S3, normalized MCD spectra measured at various magnetic fields are overlapped on one spectrum, indicating that the ferromagnetic properties come from a single ferromagnetic phase, that is the zinc-blende (Ga,Fe)Sb. Thus the ferromagnetism and the related magnetic properties such as magnetic anisotropy presented in this work are intrinsic to the zinc-blende (Ga,Fe)Sb. Figures S4(a)–S4(d) show the corresponding Arrott plots of all the $(Ga_{0.7},Fe_{0.3})Sb$ samples, which indicate that the Curie temperature $T_C$ is higher than 320 K in all the samples.

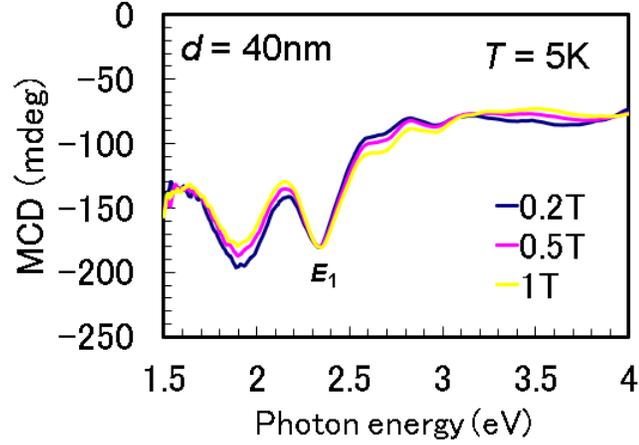

Figure S3. MCD spectra of the 40nm-thick $(Ga_{0.7},Fe_{0.3})Sb$ film measured at 5 K with a magnetic field of 0.2 T, 0.5 T, and 1 T applied perpendicular to the film plane. The MCD spectra measured at 0.2 T and 0.5 T are normalized to that at 1 T by the intensity of the $E_1$ peak.

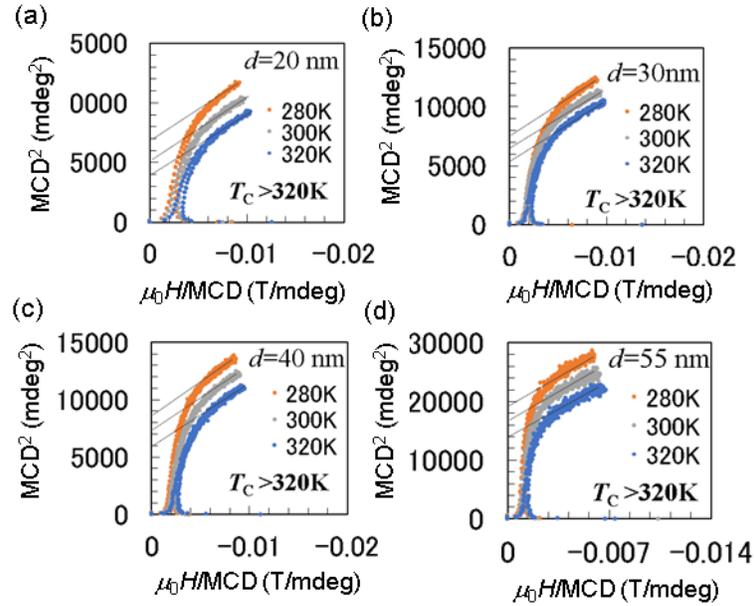

Figure S4. (a)–(d) Arrott plots $MCD^2 - H/MCD$ of $(Ga_{0.7}Fe_{0.3})Sb$ with various $d$. The (Ga,Fe)Sb thin films in all the samples exhibit clear ferromagnetism with $T_C > 320$ K.

## 3. X-ray diffraction (XRD) measurements for the estimation of epitaxial strain

In order to estimate the epitaxial strain as a function of $d$, we measured X-ray diffraction (XRD) of the (Ga,Fe)Sb samples using a Rigaku's Smart-Lab® system with a copper source at an X-ray wavelength of 0.154 nm. Figure S5 show the XRD curves of all the samples. In all the samples, by fitting Gaussian curves to the XRD peaks of the buffer layer and the (Ga,Fe)Sb layer, we determined the peak position (theta value). From the peak positions, we estimated the intrinsic lattice constants of (Ga,Fe)Sb ($a_\text{GaFeSb}$) and of the buffer layer ($a_\text{AlSb}$). We estimated the epitaxial strain $\varepsilon$ by using the following formula, $\varepsilon = \frac{a_\text{GaFeSb} - a_\text{AlSb}}{a_\text{GaFeSb}} \times 100$ (%) (see [S1] for the estimation method of $\varepsilon$). From Fig. S5, the peak positions of (Ga,Fe)Sb and AlSb are almost same in all the samples; the estimated strain value is $\varepsilon \sim -1.58$ %. Hence, all the (Ga,Fe)Sb films have tensile strain, consistent with our previous report [S1]. Therefore, we rule out any strain induced magnetic-anisotropy change in the (Ga,Fe)Sb films with various thicknesses.

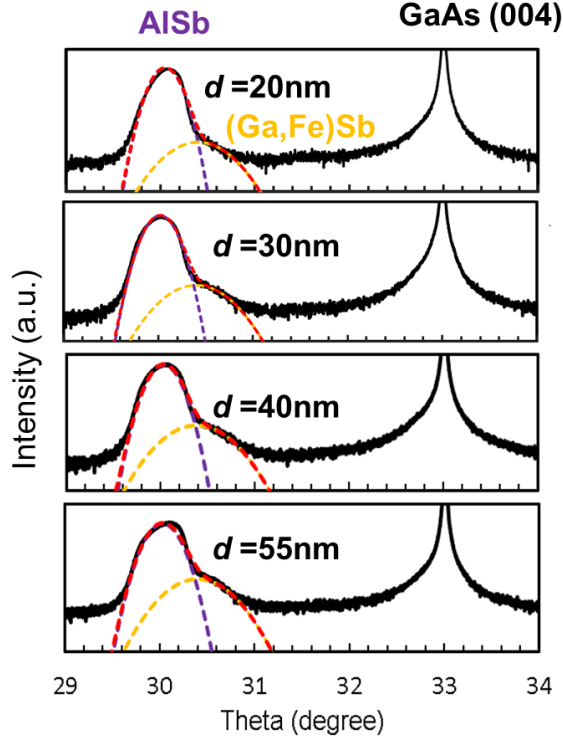

Figure S5. X-ray diffraction curves of all the $(Ga_{0.7}, Fe_{0.3})Sb$ films, with thickness $d = 20, 30, 40,$ and 55 nm, grown on an AlSb buffer layer. The broad peak (red dashed line) is fitted by the Gaussian curves corresponding to the peak of (Ga,Fe)Sb (yellow dashed line) and AlSb (purple dashed line).

## 4. Theoretical modelling of ferromagnetic resonance (FMR) spectra

(Ga,Fe)Sb is a new ferromagnetic semiconductor (FMS), and ferromagnetic resonance (FMR) has never been studied before except our recent work [S1]. However, FMR in the prototypical FMS (Ga,Mn)As has been studied extensively, and therefore in this study we use the same equations which were used to analyze the FMR results of (Ga,Mn)As. The free energy density $E$ of a (Ga,Fe)Sb thin film in the presence of an external magnetic field H is given by Ref. [S2] as shown in Eq. (S1) [same as Eq. (1) in the main text].

$$E = \frac{1}{2}\mu_0 M_S \left\{ -2H\left[\cos\theta_M \cos\theta_H + \sin\theta_M \sin\theta_H \cos(\varphi_M - \varphi_H)\right] + M_S \cos^2\theta_M - H_{2\perp}\cos^2\theta_M \right.$$

$$\left. -\frac{H_{4\perp}}{2}\cos^4\theta_M - H_{2//}\sin^2\theta_M \sin^2\left(\varphi_M - \frac{\pi}{4}\right) \right\}. \tag{S1}$$

Here in { }, the first term is the Zeeman energy, the second term is the demagnetizing energy (also called 'shape anisotropy'), and the remaining terms represent the magnetic anisotropy energy, where $\mu_0 H_{2\perp}$, $\mu_0 H_{2//}$, and $\mu_0 H_{4\perp}$ represent the perpendicular uniaxial, in-plane uniaxial, and perpendicular cubic anisotropy fields, respectively. The angles of the magnetization **M**, $\varphi_M$ (in-plane) and $\theta_M$ (out-of-plane), as well as the angles of the magnetic field **H**, $\varphi_H$ and $\theta_H$, are defined in Fig. S6. The anisotropy fields $\mu_0 H_{2\perp}$, $\mu_0 H_{4\perp}$, and $\mu_0 H_{2//}$ in Eq. (S1) are defined as $\mu_0 H_{2\perp} = 2K_{2\perp}/M_S$, $\mu_0 H_{4\perp} = 2K_{4\perp}/M_S$, and $\mu_0 H_{2//} = 2K_{2//}/M_S$, where $K_{2\perp}$, $K_{2//}$, and $K_{4\perp}$ represent the perpendicular uniaxial, in-plane uniaxial, and perpendicular cubic anisotropy constants, and $M_S$ is the saturation magnetization.

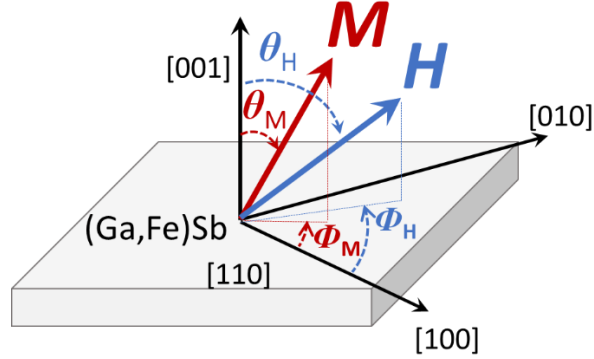

Figure S6. Coordinate system used in the modelling of FMR. The orientations of the magnetic field **H** and magnetization **M** are described by angles ($\theta_H$, $\varphi_H$) and ($\theta_M$, $\varphi_M$), respectively.

The direction of the magnetization ($\theta_M$, $\varphi_M$) at the resonant condition for an applied magnetic field $H$ direction ($\theta_H$, $\varphi_H$) can be determined by the minimization of the free energy density $E$ with respect to $\theta_M$ and $\varphi_M$.

$$\frac{\partial E}{\partial \theta_M} = 0; \tag{S2a}$$

$$\frac{\partial E}{\partial \varphi_M} = 0. \tag{S2b}$$

The resonant condition for FMR is given by the Smith-Beljers relation [S2],[S3].

$$\left(\frac{\omega}{\gamma}\right)^2 = \frac{1}{(M_S \sin \theta_M)^2} \left[\frac{\partial^2 E}{\partial \theta_M^2}\frac{\partial^2 E}{\partial \varphi_M^2} - \left(\frac{\partial^2 E}{\partial \theta_M \partial \varphi_M}\right)^2\right], \tag{S3}$$

where $\mu_0$ is the vacuum permeability, $\omega$ is the angular frequency of magnetization precession, $\gamma$ is the gyromagnetic ratio ($\gamma = g\mu_B/\hbar$), $g$ is the Landé $g$-factor, $\mu_B$ is the Bohr magneton, and $\hbar$ is the reduced Planck constant. In Eq. (S3), each term can be described as

$$\frac{\partial^2 E}{\partial \theta_M^2} = \mu_0 M_S (H_R a_1 + b_1), \quad \frac{1}{\sin^2\theta_M}\frac{\partial^2 E}{\partial \varphi_M^2} = \mu_0 M_S (H_R a_1 + b_2), \quad \frac{1}{\sin^2\theta_M}\frac{\partial^2 E}{\partial \theta_M \partial \varphi_M} = \mu_0 M_S b_3.$$

Then, Eq. (S3) is expressed as

$$\left(\frac{\omega}{\gamma}\right)^2 = \mu_0^2\left[(H_R a_1 + b_1)(H_R a_1 + b_2) - b_3^2\right], \tag{S4}$$

where

$a_1 = \cos\theta_M \cos\theta_H + \sin\theta_M \sin\theta_H \cos(\varphi_M - \varphi_H),$

$b_1 = -\left[M_S - H_{2\perp} + H_{2//}\cos^2\left(\varphi_M + \frac{\pi}{4}\right)\right]\cos 2\theta_M + H_{4\perp}\frac{\cos 2\theta_M + \cos 4\theta_M}{2},$

$$b_2 = -(M_S - H_{2\perp})\cos^2\theta_M + H_{4\perp}\cos^4\theta_M - H_{2//}\left\{\sin 2\varphi_M + \left[\cos\theta_M \cos\left(\varphi_M + \frac{\pi}{4}\right)\right]^2\right\};$$

$$b_3 = \frac{H_{2//}}{2}\cos\theta_M \cos 2\varphi_M.$$

We used Eq. (S4) to describe our experimental results.

Finally, Eq. (S3) can be rewritten for the rotation of $H$ in the perpendicular plane ($\varphi_H = \varphi_M = 45°$), *i.e.* in the ($\bar{1}10$) plane as

$$\left(\frac{\omega}{\gamma}\right)^2 = \mu_0^2 [H_R \cos(\theta_H - \theta_M) + (-M_S + H_{2\perp})\cos^2\theta_M + H_{4\perp}\cos^4\theta_M - H_{2//}]$$

$$\times \left[H_R \cos(\theta_H - \theta_M) + \left(-M_S + H_{2\perp} + \frac{H_{4\perp}}{2}\right)\cos 2\theta_M + \frac{H_{4\perp}}{2}\cos 4\theta_M\right] \quad (S5)$$

and for the rotation of $H$ in the film plane ($\theta_H = \theta_M = 90°$), *i.e.* in the (001) plane as

$$\left(\frac{\omega}{\gamma}\right)^2 = \mu_0^2\left[H_R \cos(\varphi_H - \varphi_M) + M_S - H_{2\perp} + H_{2//}\sin^2\left(\varphi_M - \frac{\pi}{4}\right)\right]$$

$$\times \left[H_R \cos(\varphi_H - \varphi_M) - H_{2//}\cos\left(2\varphi_M - \frac{\pi}{2}\right)\right]. \quad (S6)$$

Eqs. (S5) and (S6) are the same as Eqs. (3) and (4) in the main manuscript, and these are our fitting equations for our FMR results, where $\mu_0 H_{2\perp}$, $\mu_0 H_{4\perp}$, $\mu_0 H_{2//}$ and $g = (\gamma\hbar/\mu_B)$ are the fitting parameters.

**5. FMR spectra and background signal**

Figure S7 shows the FMR spectra of our $(Ga_{0.7},Fe_{0.3})Sb$ samples with $d = 20$ nm, 30 nm, 40 nm, and 55 nm, measured with and without the sample in the quartz rod at room temperature (300 K). The red curves show the raw FMR spectra of the (Ga,Fe)Sb samples, whereas the dotted black curves represent the FMR spectra without the samples. The dotted black curves are the background signals which overlapped the FMR spectra of samples. In the main manuscript, we have shown FMR data after removing these background signals.

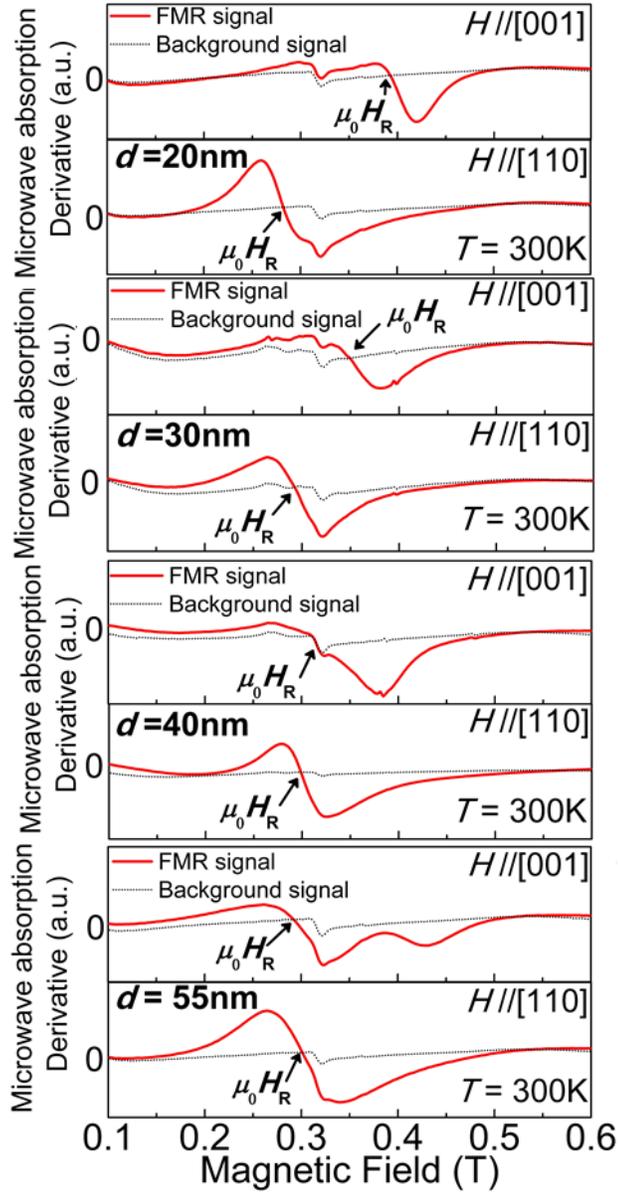

Figure S7. FMR spectra with and without the samples in the quartz rod, measured for $(Ga_{0.7}Fe_{0.3})Sb$ with d = 20 nm, 30 nm, 40 nm, and 55 nm at room temperature (300 K). The black dotted lines represent the background signals, and the red lines represent the raw FMR data with the background signals.